\documentclass[preprint,review,11pt]{elsarticle}

\usepackage{graphicx}

\usepackage{amssymb}

\usepackage[top=4cm,left=4.5cm,right=4.5cm,bottom=4.5cm]{geometry}

\usepackage{enumitem} 
  \setlist{itemsep=1ex plus0.2ex, leftmargin=*, align=left}

\makeatletter
\newcommand{\labitem}[2]{%
\def\@itemlabel{\textbf{#1}}
\item
\def\@currentlabel{#1}\label{#2}}
\makeatother

\makeatletter
\newcommand{\headingitem}[1]{%
\vspace{0.3cm}
\def\@itemlabel{\textbf{#1}}
\item
\def\@currentlabel{#1}
\addtocounter{enumi}{-1}
}
\makeatother

\usepackage{csvsimple}
\usepackage{multirow}
\usepackage{lipsum}

\usepackage{eurosym}
\usepackage{siunitx}

    \DeclareSIUnit\eur{\officialeuro}
    \DeclareSIUnit\M{M}
    \DeclareSIUnit\k{k}

  \def\sym#1{\ifmmode^{#1}\else\(^{#1}\)\fi}

\usepackage{setspace} 
\usepackage{csquotes}
\usepackage{amsmath}
\usepackage{bm}

\usepackage{xspace}
	\newcommand\ie{i.\,e.\xspace}
	\newcommand\eg{e.\,g.\xspace}

	\newcommand\cf{cf.\xspace}

	\newcommand\US{U.\,S.\xspace}

\usepackage[amsmath,hyperref,framed]{ntheorem} 
  \theoremstyle{plain}
     
  \theoremstyle{nonumberplain}
    \theoremseparator{.}
    \theoremheaderfont{\bfseries}
    \theorembodyfont{\normalfont}
    \theoremsymbol{$\blacksquare$}
      \RequirePackage{amssymb}

      \makeatletter
    \let\copy@theorem@headerfont=\theorem@headerfont
    \newcommand{\my@theorem@headerfont}{%
        \boldmath\copy@theorem@headerfont\unboldmath
      }
    \let\theorem@headerfont=\my@theorem@headerfont
      \makeatother

\theoremstyle{nonumberplain}
\theoremseparator{.}
\usepackage[%
  sort&compress
]{cleveref}
\usepackage{url}

\usepackage{isomath}

\usepackage{algorithmic}        

    \crefname{ALC@unique}{step}{steps}
    \Crefname{ALC@unique}{Step}{Steps}
    \crefname{ALC@line}{step}{steps}
    \Crefname{ALC@line}{Step}{Steps}

\usepackage[usenames,dvipsnames]{xcolor}

\usepackage{colortbl}
\usepackage{tabularx}
\usepackage{booktabs}
\usepackage{collcell}

\usepackage{ragged2e}
  
\newcommand{\PreserveBackslash}[1]{\let\temp=\\#1\let\\=\temp}
\newcolumntype{v}[1]{>{\PreserveBackslash\RaggedRight\hspace{0pt}}p{#1}}

  \usepackage{adjustbox}
\newcolumntype{Q}[2]{%
    >{\adjustbox{angle=#1,lap=\width-(#2)}\bgroup}%
    l%
    <{\egroup}%
}

\newcommand{\mcellt}[2][c]{%
  \begin{tabular}[t]{@{}#1@{}}#2\end{tabular}}

\usepackage[
    colorinlistoftodos,
    textsize=footnotesize,
        ]{todonotes}

\makeatletter
    \renewcommand{\fps@figure}{htb}         
    \renewcommand{\fps@table}{htb}         
\makeatother 

\usepackage{float}
\usepackage{rotating}

\journal{Decision Support Systems}

\begin{document}

\begin{frontmatter}



\title{News-based trading strategies}


\author[Freiburg,NII]{Stefan Feuerriegel\corref{cor1}}
\ead{stefan.feuerriegel@is.uni-freiburg.de}

\author[NII]{Helmut Prendinger}
\ead{helmut@nii.ac.jp}

\address[Freiburg]{Chair for Information Systems Research, University of Freiburg, Platz der Alten Synagoge, 79098 Freiburg, Germany}
\address[NII]{National Institute of Informatics~(NII), 2-1-2 Hitotsubashi, Chiyoda-ku, Tokyo 101-8430, Japan}
\cortext[cor1]{Corresponding author. Mail: stefan.feuerriegel@is.uni-freiburg.de; Tel: +49\,761\,203\,2395; Fax: +49\,761\,203\,2416.}

\begin{abstract}
The marvel of markets lies in the fact that dispersed information is instantaneously processed and used to adjust the price of goods, services and assets. Financial markets are particularly efficient when it comes to processing information; such information is typically embedded in textual news that is then interpreted by investors. Quite recently, researchers have started to automatically determine news sentiment in order to explain stock price movements. Interestingly, this so-called news sentiment works fairly well in explaining stock returns. {In this paper, we design trading strategies that utilize textual news in order to obtain profits on the basis of novel information entering the market. We thus propose approaches for automated decision-making based on supervised and reinforcement learning.} {Altogether, we demonstrate how news-based data can be incorporated into an investment system.} 
\end{abstract}


\end{frontmatter}



\section{Introduction}

Market efficiency relies, to a large extent, upon the availability of information. Nowadays, market information can be accessed easily as it comes na{\"i}vely with the prevalence of electronic markets. Then, decision-makers can use such information to maximize the benefit of purchases and sales (\eg~\cite{Granados.2010}). 
 In the same context, several publications (\eg~\cite{Cenesizoglu.2014,Cutler.1989,Overshooting.APE}) study the market reception of news announcements, finding a causal and clearly measurable relationship between financial disclosures and stock market reaction. Market reception is not only shaped by the quantitative facts embedded in financial disclosures, but, more importantly, stock market reactions to financial disclosures are driven by qualitative information, since news is typically embodied in text messages. In order to extract tone from textual content, one frequently measures the polarity of news by measuring the so-called \emph{news sentiment}. 
This demonstrates how the narrative content of disclosures~\cite{Balakrishnan.2010,Chan.2011,Fersini.2014,Li.2013,Negations.DSS,Wang.2014} can be harnessed to provide \emph{decision support} for investors.

While previous research~\cite{Antweiler.2004,Tetlock.2007,Tetlock.2008,Henry.2008} succeeded in establishing a link between news tone and stock market prices, it was not clear how the extracted sentiment signals could then be utilized to facilitate investment decisions. To close this gap, this paper studies how news sentiment, as a recent trend of decision support, can enrich news-based trading. News trading combines real-time market data and natural language processing to detect suitable news announcements in order to trigger transactions. Its mechanisms are often part of an algorithmic trading system, while many regard it as an enabling Decision Support System~(DSS) for use in banking and financial markets~\cite{Gagnon.2013}. 

With the increasing statistical reliability of text mining algorithms, news vendors now actively integrate this technology into their traditional platforms. For example, Thomson Reuters offers additional information along with their news products, such as Thomson Reuters News Analytics~(TRNA) scores, which measure the polarity and novelty of news content. The reasons for this development predominantly result~\cite{Gagnon.2013} from recent advances in natural language processing, coupled with cheaper computational power and more diversified news sources. Consequently, users are motivated to harness such a Decision Support System due to its direct gains, based on the relative value added or the advantage of such a system compared to existing approaches~\cite{Gagnon.2013}. 

Consequently, this paper investigates how a Decision Support System can utilize news sentiment to perform stock trading in practice. 
Several papers~\cite{Schumaker.2009,Schumaker.2009b,Schumaker.2012} from the DSS domain elaborate on a general system design, but do not go as far as comparing approaches to trading signals and evaluating the accuracy of the resulting decisions within a financial portfolio. 
Overall, our contribution is as follows: first, we implement different rule-based trading strategies and propose the use of automated learning strategies. Second, we find quantitative evidence that our news trading system can successfully incorporate news-based data in order to make investment decision. In addition, news-based trading benefits from incorporating other external variables, such as price momentum, to achieve higher profits. 

The scope of this Decision Support System goes beyond the scenario of news sentiment, since it can be utilized in any situation where novel information enters the market and triggers a subsequent response. Examples include any firm-related information, such as press releases, earnings calls and sales figures. Accordingly, the DSS itself does not unveil hidden patterns to generate excess profits. On the contrary, it relies purely on novel information entering the market causing an adjustment of stock prices. Hence, the profits are not the result of arbitrage \cite{McLean.2016} and unlikely to diminish to zero-excess returns in the future according to the semi-strong form of the efficient market hypothesis \cite{Fama.1965}. 

The remainder of this paper is structured as follows. In Section~\ref{sec:related_work}, we review related research on news trading. Next, Section~\ref{sec:background} describes the data sources, as well as the news corpus, that are integrated into the sentiment analysis to extract the subjective tone of financial disclosures. The calculated sentiment values are then inserted (Section~\ref{sec:trading_strategies}) into various news trading strategies and, finally, Section~\ref{sec:evaluation} evaluates these strategies in terms of their financial performance. 

\section{Related work}
\label{sec:related_work}

This section introduces a short description of components that are relevant for a news trading system. For a thorough review and taxonomy, we refer to the reference~\cite{Gagnon.2013}. 

\subsection{Benchmarks}


First of all, we need to validate the performance of our trading strategies using benchmark scenarios. Among the prevalent approaches is the use of stock indices, as these aggregate individual stock investments weighted by size. For example, previous research predicts the direction of stock price movements via sparse matrix factorization based on news stories from the \emph{Wall Street Journal}~\cite{Wong.2014}. Their findings reveal an accuracy of \SI{55.7}{\percent}, which exceeds their reference index. Further alternatives build on simple buy-and-hold strategies of stocks with the highest historic returns. 

\subsection{Transaction fees}

When aiming for a realistic study, another integral part of news trading that we need to account for is incurred \emph{transaction fees}. Although prior research has conducted trading simulations, many of these references neglect the influence of transaction fees.

{According to~\cite{Graf.2011}, trading strategies based on positive and negative sentiment can be profitable if the transaction costs are moderate. The paper assumes an investor who trades in \num{62}~stocks simultaneously based on Reuters company news. In addition, the author considers transaction costs and possible losses from interest rates when no endowment is considered. The findings reveal that trades on sell-signals are more profitable but less frequent. Here, sensitivity is studied by varying transaction costs and risk-free interest rates but the author does not evaluate any trading strategies, excepting the na{\"i}ve strategy.}

{A straightforward approach relies on classification rules based on risk words~\cite{Li.2006}, resulting in an average annual excess of \SI{20}{\percent} compared to \US Treasury bills. Similarly, Tetlock~\cite{Tetlock.2007} utilizes pessimistic words and obtains \SI{7.3}{\percent} higher returns, when compared to the Dow Jones, with the applied trading strategy. However, all the aforementioned papers ignore transaction fees, which, in fact, can be substantial~\cite{Hagenau.2012b}.}

{As another example, a support vector machine using financial news achieves an accuracy of \SI{71}{\percent} in predicting the direction of asset returns~\cite{Schumaker.2009}. Hence, this creates an excess return of \SI{2.88}{\percent} compared to the S\&P~500 index between October~25, 2005 and November~28, 2005.}

{One of the few papers that considers transaction fees also utilizes German ad~hoc announcements and achieves an accuracy of \SI{65}{\percent} when predicting the direction of returns. The average return per transaction accounts for \SI{1.1}{\percent} when assuming transaction fees of \SI{0.1}{\percent}. However, this paper~\cite{Hagenau.2012b} relies on one basic strategy (similar to our simple news-based strategy) and does not compare other trading strategies.}

\subsection{Trading strategies}

The third component of a news trading system entails trading strategies. Common approaches include simple buy-and-hold strategies which are equipped with information from news~\cite{LNBIP.NewsTrading}. The trading strategies are then usually tested in an ex post portfolio simulation. For instance, previous research hypothesizes that a sentiment-based selection strategy will outperform a simple buy-and-hold benchmark strategy, which holds all stocks over the whole test period~\cite{Klein.2011}. Another research stream develops a news categorization and trading system to predict stock price trends~\cite{Mittermayer.2004}. Its trading engine recommends trades, such as \emph{\textquote{buy stock $X$ and hold it until the stock prices hit the $+d$\,\% barrier}}. Similarly, a trading strategy can be built around social media data~\cite{Hochreiter.2015} or the Google query volume~\cite{Preis.2013} for search terms related to finance. The latter variable is inserted into a simple buy-and-hold strategy (without transaction costs) to buy the Dow Jones index at the beginning and sell it at the end of various holding periods. This approach yields \SI{16}{\percent} profit, almost identical to the total gains of the Dow Jones index in the time period from January 2004 until February 2011. 

A recent research paper~\cite{Takeuchi.2013} utilizes deep learning to train an autoencoder of stacked restricted Boltzmann machines~(RBM) to extract features from stock movements. These are then passed into a neural network to classify future performances. This approach yields a higher return when compared to a basic momentum trading strategy; however, it neglects the possible predictive power of financial disclosures.

Related to our research is~\cite{Gagnon.2013}, who propose an architecture for a rule-based news trading system. This system screens events with the help of different data mining algorithms and recommends trading strategies. The author thus provides a taxonomy of news trading; however, the mechanisms behind the systems are not empirically evaluated.

All in all, the above references provide evidence that studying trading strategies, with a particular focus on news content, is both an intriguing and relevant research question for the decision support domain. 
 Thus, this paper aims to shed light on this research area by comparing different trading strategies in terms of their financial performance. 

\section{Background}
\label{sec:background}

This section introduces background knowledge for both datasets and the sentiment analysis. First, we describe the construction of the news corpus that is used throughout this paper. We then transform this running text into machine-readable tokens to measure news sentiment.\footnote{See the online appendix for a specification of our preprocessing.} 

\subsection{Data sources}

Our news corpus originates from regulated ad~hoc announcements\footnote{Kindly provided by Deutsche Gesellschaft f{\"u}r Adhoc Publizit{\"a}t (DGAP).} in English. We choose this data source primarily because companies are bound to disclose these ad~hoc announcements as soon as possible through standardized channels, thereby enabling us to study the short-term effect of news disclosures on stock prices. Ad~hoc announcements are a frequent choice~\cite{Mittermayer.2006b} when it comes to evaluating and comparing methods for sentiment analysis. In addition, this type of news corpus offers several advantages: ad~hoc announcements must be authorized by company executives, the content is quality-checked by the Federal Financial Supervisory Authority\footnote{Bundesanstalt f{\"u}r Finanzdienstleistungsaufsicht (BaFin).} and several publications analyze their relevance to the stock market -- finding a direct relationship (\eg~\cite{Muntermann.2007}). Our collected announcements date from January~2004 until the end of June~2011 (due to data availability). We investigate such a long time period to avoid the possibility of only analyzing news driven predominantly by a single market regime -- for example, the financial crisis. 

We later also integrate a stock market index into our analysis as follows: in our analysis, the so-called CDAX index works as a benchmark which the trading strategies need to surpass in terms of performance.\footnote{The CDAX 
is a German stock market index calculated by Deutsche B{\"o}rse. It is a composite index of all stocks traded on the Frankfurt Stock Exchange that are listed in the General Standard or Prime Standard market segments, currently giving a total of 331~stocks.}


\subsection{Sentiment analysis}

Methods that use the textual representation of documents to measure positive and negative content are referred to as opinion mining or \emph{sentiment analysis}~\cite{Pang.2008}. In fact, sentiment analysis can be utilized~\cite{Mittermayer.2006b,Minev.2012} to extract subjective information from text sources, as well as to measure how market participants perceive and react to news. One uses the observed stock price reactions following a news announcement to validate the accuracy of the sentiment analysis routines. Thus, sentiment analysis provides an effective tool to study the relationship between news content and its market reception~\cite{Antweiler.2004,Tetlock.2007}. Let the variable $S(A)$ give the news sentiment of an announcement $A$ corresponding to stock $i$.  All algorithmic details are provided in the online appendix. 

\section{Trading strategies}
\label{sec:trading_strategies}

\begin{figure}
\centering
\includegraphics[width=.99\linewidth]{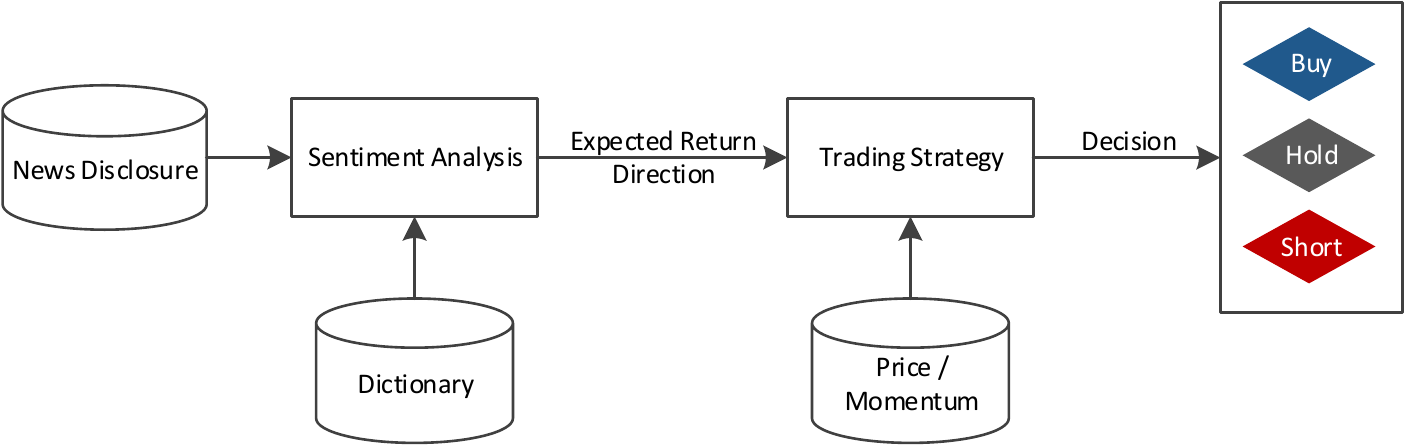}
\caption{{Decision Support System~(DSS) transforms news content into trading decisions.}}
\label{fig:research_model}
\end{figure}

{The generic design of a decision support system for news trading is depicted in Fig.~\ref{fig:research_model}. The system's internal process contains two steps: first, the system extracts the news sentiment as a measure of textual tone. This provides an indicator of the expected return direction. Then, the second step involves executing a trading strategy in order to render a trading decision.}

This section introduces the trading strategies that serve as a foundation for our analysis. Consistent with the existing literature, we start by presenting our benchmarks, namely, a momentum trading and a portfolio approach. These strategies derive purchase decisions solely from the historic returns of assets by maximizing the so-called rate-of-change (see \cite{LNBIP.NewsTrading} for details). In addition, we propose news-based trading strategies in which investment decisions are triggered by news sentiment signals. Then, we combine both methods and develop a strategy that utilizes both historic prices and news sentiment. Finally, we utilize supervised and reinforcement learning for automated learning of such rules. 

When trading, we exclude all so-called penny stocks (\ie stocks below \EUR{5}) from our evaluation
. The reason behind this is that these penny stocks tend to react more unsystematically to trends and news announcements and, consequently, may introduce a larger noise component in our data.

\subsection{Benchmarks: momentum trading and portfolio approach}

In the subsequent algorithms, we use the following notation: let $p_{i,t}$ denote the closing price of a stock $i$ at time $t$. Past stock returns can be a predictor of future firm performance. This is why we use the rate-of-change
\begin{equation}
RoC_{i,t} = \frac{p_{i,t} - p_{i,t-\delta}}{p_{i,t-\delta}} .
\end{equation}
as one of our benchmarks. In that case, we always pick the stock with the highest absolute value in terms of rate-of-change and place a corresponding buy or sell decision. In addition, we also insert the rate-of-change into a portfolio approach of 20~stocks. That is, we always select the a uniform share across the stocks with the highest rate-of-change and buy/short-sell stocks accordingly. As a result, we average the returns but also the risk across several stocks. A detailed motivation, as well as a full specification, are given in the online appendix.

\subsection{Rule-based news trading}

We now focus on news sentiment in order to enable news-based purchase decisions. In order to react to news sentiment signals, our Decision Support System needs to continuously scan the news stream and compute the sentiment once a new financial disclosure is released. When the news sentiment associated with this press release is either extremely positive or negative, this implies a strong likelihood of a subsequent stock market reaction in the same direction. We benefit from the stock market reaction if an automated transaction is triggered shortly before the price adjustment. 

To achieve this goal, we specify the so-called simple \emph{news trading strategy} (the pseudocode given in the online appendix). It triggers buy and short-sell decisions, whenever the absolute value of the news sentiment metric of an incoming announcement exceeds a certain positive or negative threshold. This decision is given as a condition, \ie if $S(A)$ is smaller than a negative threshold $\theta_S^-$ or larger than a positive $\theta_S^+$. We choose suitable threshold values for both $\theta_S^-$ and $\theta_S^+$ as part of our evaluation in \Cref{sec:evaluation}. 


%

\subsection{Combined strategy with news and momentum trading}

The subsequent trading strategy combines the above approaches by utilizing both news sentiment and historic prices in the form of momentum. We develop this trading strategy around the idea that we want to invest in assets with both (1)~a news disclosure with a high polarity and (2)~previous momentum in the same direction. Only if both the news release and historic prices give an indication of a development in the same direction does the strategy trigger a corresponding trading decision (\cf online appendix for the pseudocode).




\subsection{Strategy learning}

{Rule-based algorithms lack the flexibility to adapt to arbitrary patterns. As a remedy, supervised learning features such adaptability, as it can learn patterns from data and incorporate this to improve its predictive performance. In addition, machine learning algorithms can usually handle non-linearity as another benefit. In contrast to rules, they can be automatically calibrated by mathematical means, often have more degrees-of-freedom than the above rules and thus might better adapt to the specific task.}

{As our default approach from machine learning, we utilize random forests, as these perform well out-of-the-box \cite{Breiman.2001}. Due to these favorable characteristics, we have decided upon random forest, but we note that they can be replaced by any other supervised learning strategy. In fact, our goal is merely to show the advantages of machine learning itself.}\footnote{{We thus regard random forests as a generic prediction model that is likely to be improved (or changed) in practical applications.}}

In addition, we observe a potential dilemma: each trading decision influences the next; \ie having bought a specific stock makes the next decision -- selling or holding -- dependent on the trading history. As a remedy, we propose the use of reinforcement learning, which allows for the learning of a suitable trading sequence directly through trial-and-error experience~\cite{Sutton.1998}. In order to store current knowledge, the reinforcement learning method introduces a so-called \emph{state-action function} $Q(s_t,a_t)$ that defines the expected value of each possible action $a_t$ in each state $s_t$. If $Q(s_t,a_t)$ is known, then the optimal policy $\pi^*(s_t,a_t)$ is given by the action $a_t$ that maximizes $Q(s_t,a_t)$ given the state $s_t$. Here, we use the so-called \emph{Q-Learning} method~\cite{Sutton.1998}. We initialize the action-value function $Q(s,a)$ to zero for all states and actions. Subsequently, the agent successively observes a sequence of ups and downs from a historic dataset.

In the following, we provide details on the specification of the learning. In case of the random forest, we use the optimal decisions based on a pre-calculation for each news disclosure as a training dataset. These are a categorical variable for holding the existing stock, as well as investing in a new stock or the index. Input variables to the random forest are then the continuous values of the sentiment, the rate-of-change of the stock, the rate-of-change of the index, the historic performance of the index and a penny stock dummy, all for the current and previous news disclosure. In the case of reinforcement learning, we use two states representing the current and previous disclosure. Each encodes a combination of binary variables denoting the sign of the sentiment, the rate-of-change, the rate-of-change of the index and the historic performance of the index.  


\section{Evaluation}
\label{sec:evaluation}

The above sections have presented a number of trading strategies that differ in the way in which operations are derived; this section evaluates these trading strategies in terms of their achieved performance. We first focus on our benchmark strategies and then analyze their performance.



\subsection{Benchmarks: stock market index and momentum trading}

As our first benchmark, we choose the so-called CDAX, a German stock market index calculated by Deutsche B{\"o}rse. It is a composite index of all stocks traded on the Frankfurt Stock Exchange that are listed in the General Standard or Prime Standard market segments. During the period from 2004 until mid-2011, the index increased by \SI{50.99}{\percent}, which corresponds to \SI{5.65}{\percent} at an annualized rate. The number of days with positive returns outweighs the negative by 1,092 to 864.  

\begin{figure}
\centering
\hspace*{-2cm}
\includegraphics[width=.45\linewidth]{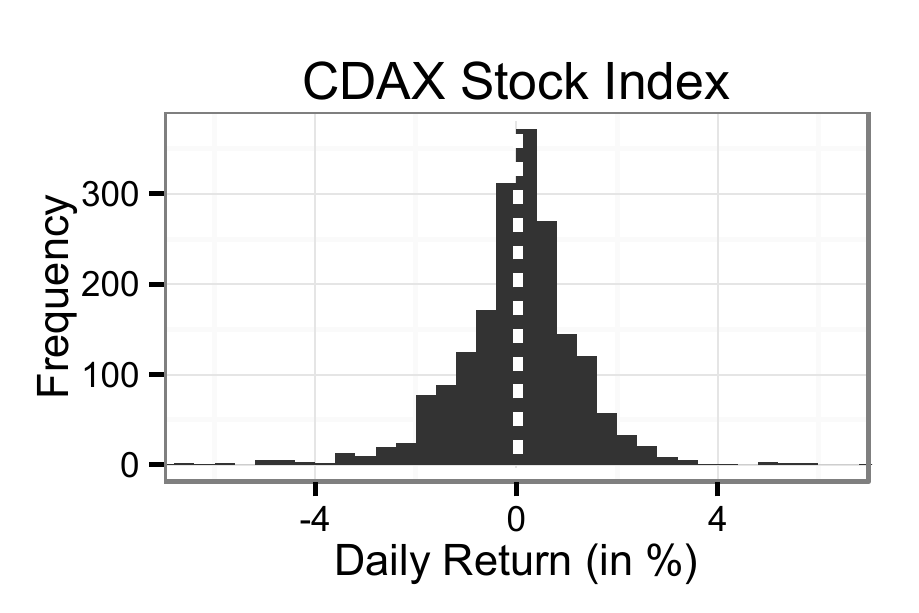}~\includegraphics[width=.45\linewidth]{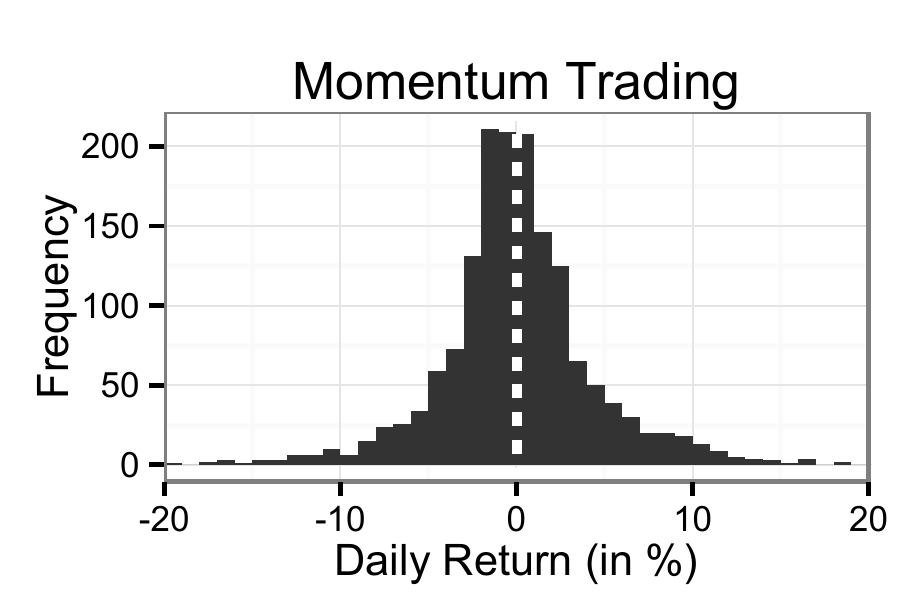}~\includegraphics[width=.45\linewidth]{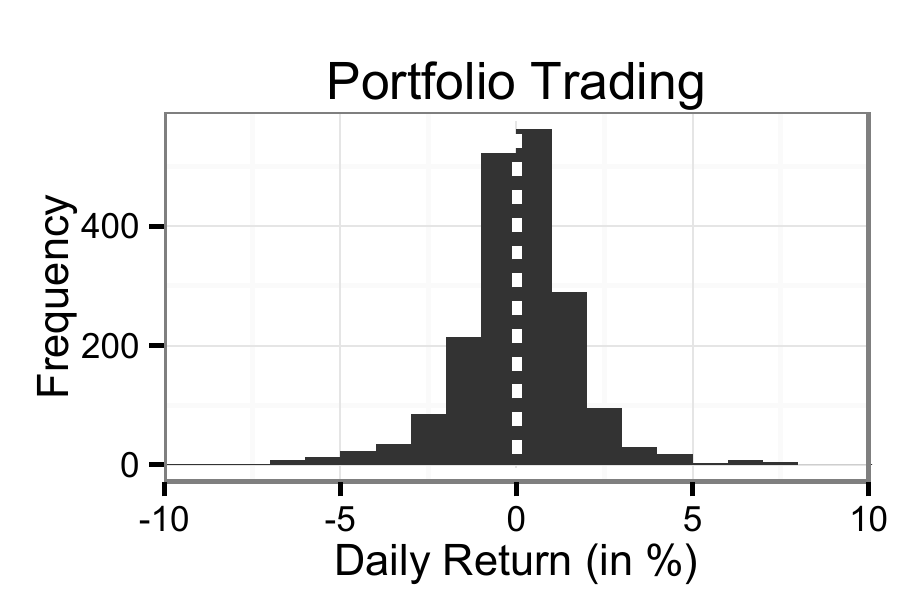}
\caption{Histogram of non-zero returns for the CDAX index~(left), momentum trading~(middle) and portfolio trading~(right), in which the vertical bars denote the corresponding mean value.}
\label{fig:cmp_cdax_momentum_trading_histogram}
\end{figure}

Simple momentum trading acts as our second benchmark. This strategy works with no data input other than historic stock prices. When historic prices continue their trend, we can invest in the specific stock to profit from this development. In addition, portfolio trading aggregates momentum trading for several stocks in order to reduce the risk component, but this also lowers the average daily returns to \SI{0.0220}{\percent}.
 When looking at the histogram of returns in Fig.~\ref{fig:cmp_cdax_momentum_trading_histogram}, we see that volatility is smaller for the CDAX. The volatility of daily returns stands at \num{0.013} for the CDAX index, while it is higher at \num{0.020} in the case of portfolio trading and at \num{0.045} in the case of momentum trading.

\subsection{News trading}

This section evaluates different variants of news trading, starting with a simple trading strategy. This strategy triggers transactions whenever a very positive or negative news release is disclosed.\footnote{Here, we take into account only the first news release of each day and only on business days, giving a total corpus of 1,892~disclosures.} {Thereby, the goal is to demonstrate how news-based data can be incorporated into an investment decision model.}

\begin{figure}
\centering
\includegraphics[width=.6\linewidth]{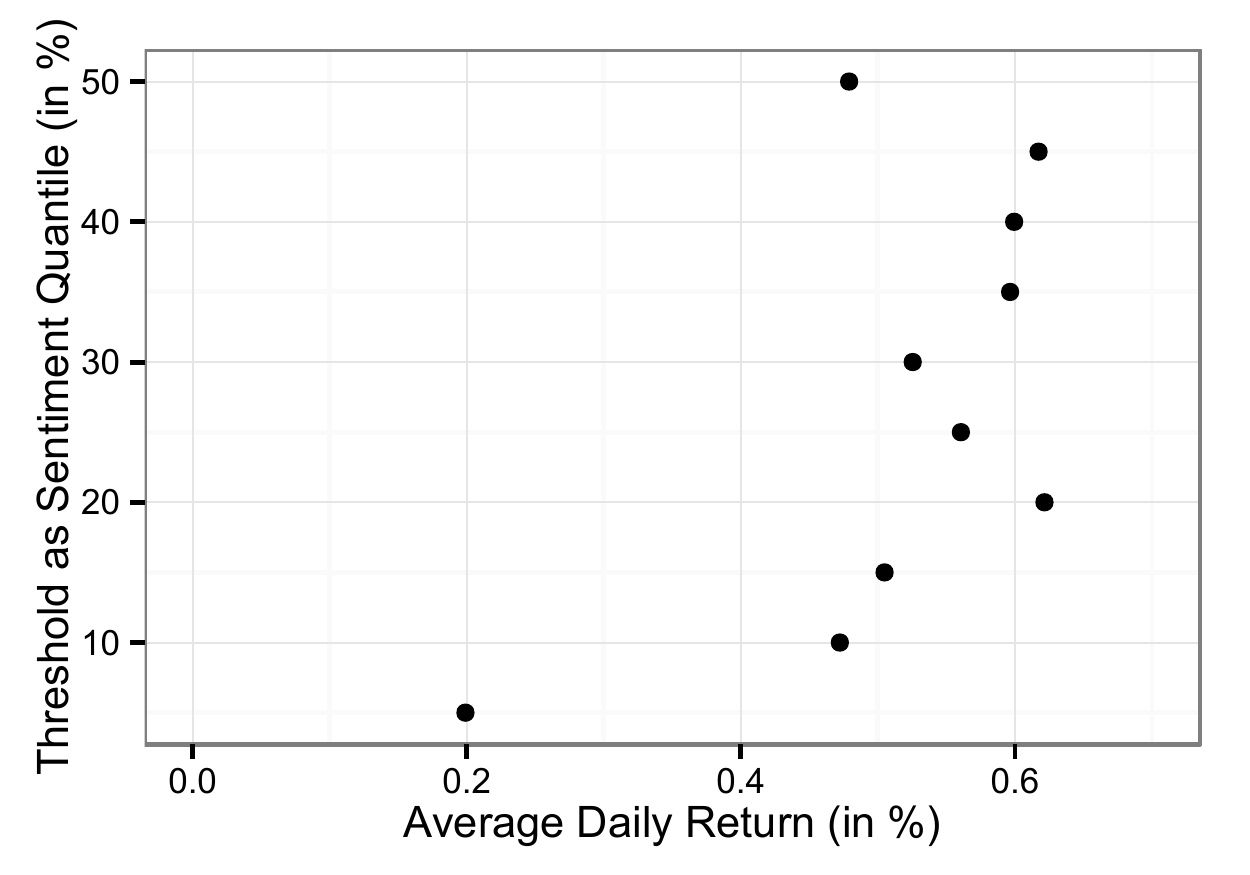}
\caption{Comparison of threshold $\theta_S^-$ (and $\theta_S^+ := \SI{100}{\percent} - \theta_S^-$ against average daily returns. The threshold is measured as quantiles from both ends of the average news sentiment in the corpus.}
\label{fig:grid_news_trading}
\end{figure}

What remains unanswered thus far is the value for the threshold $\theta_S$ above which our news-based trading strategies carry out a purchase decision. In order to find the optimal parameter, Fig.~\ref{fig:grid_news_trading} compares the thresholds $\theta_S^+$ and $\theta_S^-$ against the average returns. For reasons of simplicity, we measure these thresholds in terms of quantiles of the news sentiment distribution. We see that a threshold value of around \SI{10}{\percent} appears in a cluster of high daily returns and yields good results. Thus, we decided to set $\theta_S^-$ to the \SI{10}{\percent} quantile (and $\theta_S^+$ to the \SI{90}{\percent} quantile) of the sentiment values $S(A)$ in order to make this variable exogenously given. However, it is important to stress that there are large variations in performance depending on the threshold. 

Evaluating the above strategies with historic data reveals the following findings: with the threshold set to the \SI{10}{\percent} quantile, we gain average daily returns of \SI{0.4722}{\percent}
 at a volatility of \num{0.078}. 
  In addition, we include a combination of news and momentum trading. This strategy leads to a lower performance with 
 average daily abnormal returns of \SI{0.0335}{\percent}. However, this strategy simultaneously reveals a reduced risk component in the form of less volatility, which stands at \num{0.028}. 

\begin{figure}
\centering
\includegraphics[width=.8\linewidth]{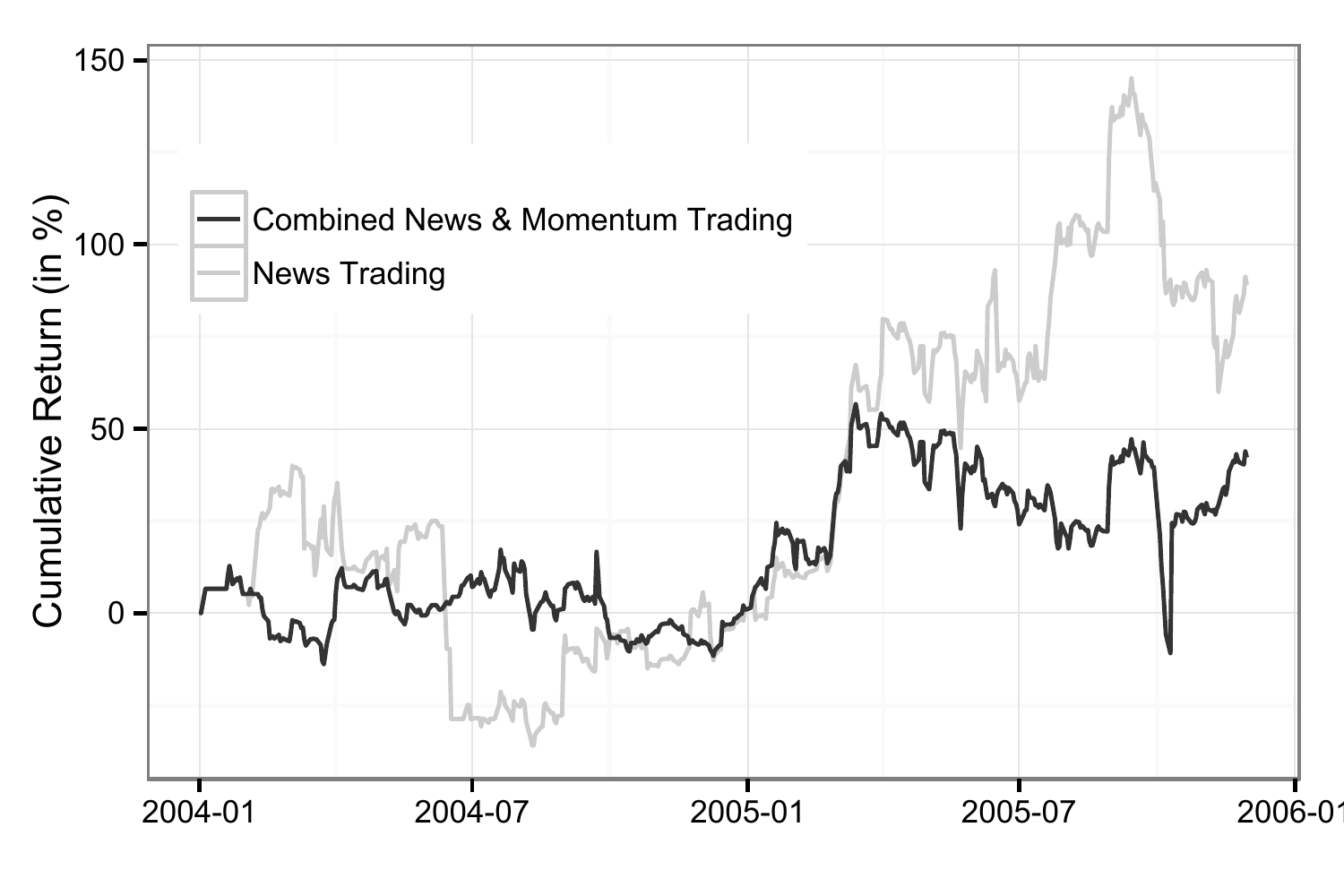}
\caption{Cumulative returns of both news trading and the combination of news and momentum trading compared across the first \num{500}~business days.}
\label{fig:cmp_news_trading}
\end{figure}

\begin{figure}
\centering
\includegraphics[width=.45\linewidth]{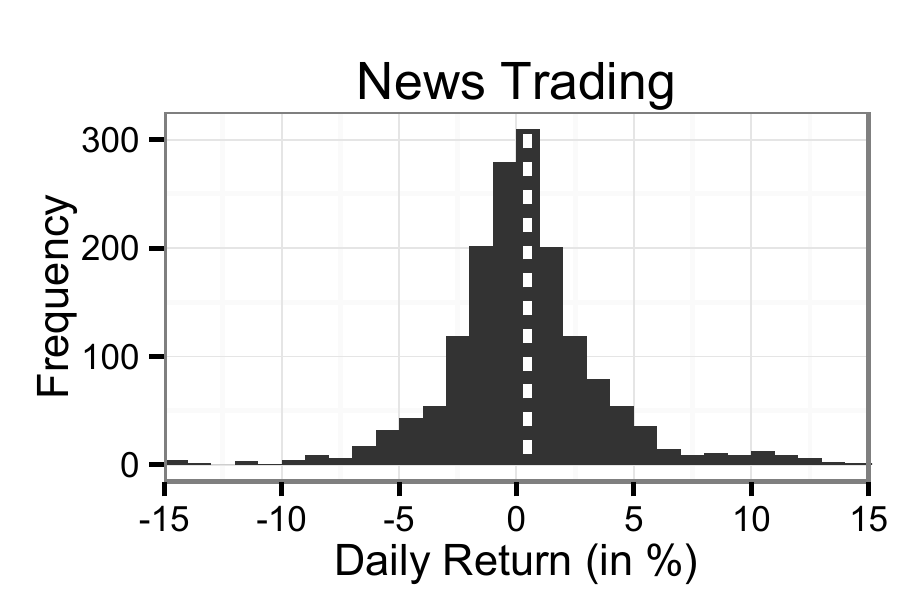}
\includegraphics[width=.45\linewidth]{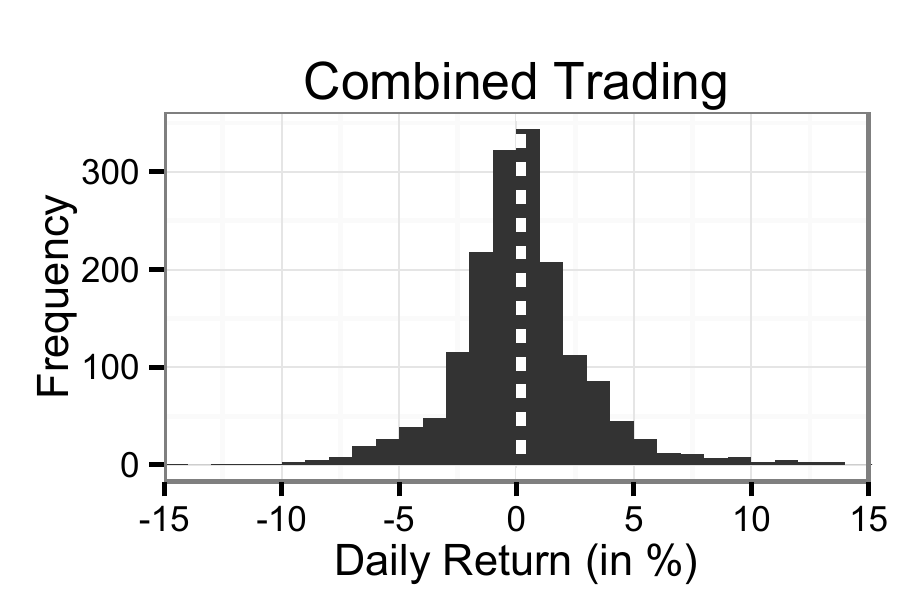}
\caption{Histogram of non-zero returns comparing simple news trading~(left) and the combined news \& momentum trading strategy~(right), in which the vertical bars denote the corresponding mean value.}
\label{fig:cmp_news_trading_histogram}
\end{figure}

Both strategies -- namely, simple news trading and the combined version -- are further evaluated in the following diagrams. {Fig.~\ref{fig:cmp_news_trading} depicts how the cumulative returns develop during the first \num{500}~business days. In other words, this figure shows how the value of an investment portfolio evolves over time when starting with one monetary unit. After a highly volatile beginning (also with negative valuations), the values of both portfolios increase considerably after around one year. The rise is more substantial in the case of simple news trading in comparison to the combined strategy.} Analogously, Fig.~\ref{fig:cmp_news_trading_histogram} compares the distribution of daily returns. We note an evidently higher volatility in the case of simple news trading, which is ultimately linked to higher risks. 

\begin{figure}
\centering
\includegraphics[width=.45\linewidth]{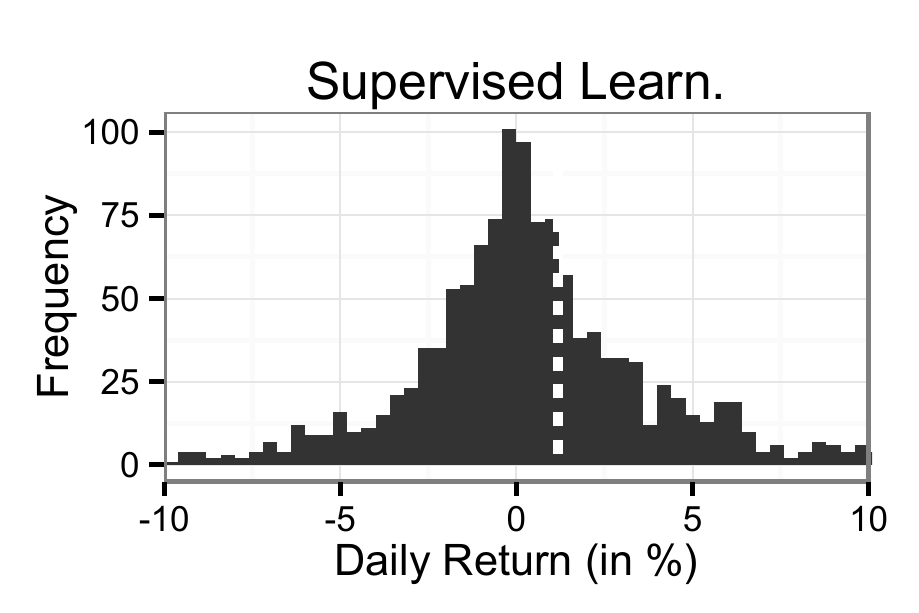}
\includegraphics[width=.45\linewidth]{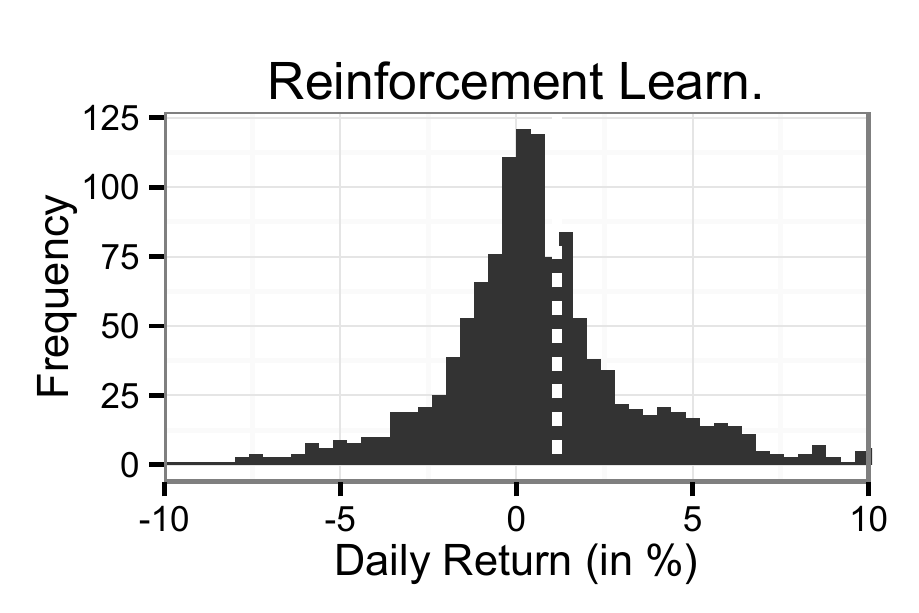}
\caption{Histogram of non-zero returns for the supervised learning in form of a random forest~(left) and the reinforcement learning~(right), in which the vertical bars denote the corresponding mean value (evaluated from years 2006 onwards, while trained with years 2004 and 2005).}
\label{fig:cmp_strategy_learning_trading_histogram}
\end{figure}

Both methods for strategy learning perform more trades than other news trading strategies, \ie a trade happens every second to third business day. Nevertheless, both achieve among the highest average returns. 
 While returns are more favorable for the random forest as a supervised learning approach, we face a lower risk with reinforcement learning. We also note that rule-based news trading features more days with profitable trades, though these do not necessarily translate into financial gains. Apparently, strategy learning is able to identify those days with higher returns, as is also shown in the histograms in Fig.~\ref{fig:cmp_strategy_learning_trading_histogram}.

\subsection{Comparison}

The simulation horizon starts in January~2004 and then spans a total of 1,956~business days. All results of our trading simulation are provided in Table~\ref{tbl:comparison}. Here, we evaluate how well the investment decisions of each strategy accord with market feedback. We focus mainly on average daily return, since cumulative returns can be misleading. The reasoning is as follows: one wrong trade can cause performance to plummet, while a high average daily return indicates sustained benefit. In addition, we want to direct attention to the volatility column. These values serve as an indicator of the level of risk associated with each strategy. Even though simple news trading achieves higher returns, it is linked with higher volatility and higher risks. Thus, it may be beneficial for practitioners to follow a strategy that results in smaller returns, while also decreasing the associated risk. In addition, we report abnormal returns which correct for concurrent market returns (\cf the online appendix for the corresponding calculation). Median returns appear to be \SI{0}{\percent} as prices for certain stocks might not change every day due to liquidity issues or a high ratio of index trades in case of median abnormal returns.

{We now compare our benchmarks to news trading in order to demonstrate how news-based data can be incorporated into an investment decision model.} The benchmarks feature mean returns of \SI{0.0298}{\percent} for the CDAX and \SI{0.0464}{\percent} for momentum trading. In comparison, news trading reaches \SI{0.4722}{\percent} for the simple strategy and \SI{1.1807}{\percent} in the supervised case. This is a visible increase, but linked with a considerably higher volatility. Interestingly, we can diminish the risk component by looking at specific sectors, such as automotive or chemicals. In both cases, we see a reduction in volatility at the cost of decreasing mean returns. Furthermore, we note that the inherent prediction errors of machine learning algorithms add a further source of noise and thus impairs the risk component compared to rules. We stress that this point that we work with daily returns, while results might differ for intraday returns. As a remedy, we provide a comparison with abnormal returns in \Cref{tbl:comparison}. In addition, \Cref{sec:limitations} discusses generalizability and limitations. 

\begin{table}[htbp]
\caption{Comparison of benchmarks and trading strategies across several key performance characteristics. Median returns can be \SI{0}{\percent} as prices might not change every day due to liquidity issues or a high ratio of index trades. The Sharpe ratio is calculated in regard to the CDAX index.}
\label{tbl:comparison}
\centering
\scriptsize
\hspace*{-2cm}
\rotatebox{90}{%
\begin{tabular}{l S SSSS S}
\toprule
\textbf{Trading strategy} &
\textbf{\mcellt{Returns:\\ mean}} &
\textbf{\mcellt{Returns:\\ median}} &
\textbf{\mcellt{\#Positive\\ returns}} & 
\textbf{\mcellt{\#Negative\\ returns}} & 
\textbf{\mcellt{\#Total\\ trades}} &
\textbf{$\Delta$Trades}  \\
\midrule
\textbf{Benchmarks} \\
CDAX index & \SI{0.0298}{\percent} & \SI{0.0703}{\percent} & 1092 & 864 & 1 & {---} \\
Momentum trading & \SI{0.0464}{\percent} & \SI{0.0000}{\percent} & 1129 & 827 & 19 & \SI{102.95}{d}\\
\multicolumn{5}{l}{($\theta = \SI{50}{\percent}$, $\delta = \SI{200}{d}$)} \\
Portfolio trading (20 stocks) & \SI{0.0220}{\percent} & \SI{0.0593}{\percent} & 1044 & 912 & 1956 & \SI{1.00}{d} \\
\midrule
\textbf{News trading} \\
Simple news trading & \SI{0.4722}{\percent} & \SI{0.0000}{\percent} & 1164 & 792 & 196 & \SI{9.98}{d} \\
\multicolumn{5}{l}{($\theta_S^+ = \SI{90}{\percent}$, $\theta_S^- = \SI{10}{\percent}$)} \\
Sector-specific: automobile & \SI{0.0116}{\percent} & \SI{0.0000}{\percent} & 1125 & 831 & 20 & \SI{97.80}{d} \\
Sector-specific: chemicals & \SI{0.0184}{\percent} & \SI{0.0000}{\percent} & 1105 & 851 & 5 & \SI{391.20}{d} \\
Combined: news \& momentum & \SI{0.2001}{\percent} & \SI{0.0000}{\percent} & 1142 & 814 & 99 & \SI{19.76}{d} \\
\multicolumn{5}{l}{($\theta_S^+ = \SI{90}{\percent}$, $\theta_S^- = \SI{10}{\percent}$, $\delta = \SI{200}{d}$)} \\
\midrule
\multicolumn{5}{l}{\textbf{Strategy learning}$^\dagger$} \\
Supervised learning & \SI{1.1807}{\percent} & \SI{0.0000}{\percent} & 835 & 598 & 589 & \SI{2.43}{d} \\
\multicolumn{5}{l}{(Random forest with $n=500$ trees)} \\
Reinforcement learning & \SI{1.1506}{\percent} & \SI{0.2549}{\percent} & 910 & 523 & 480 & \SI{2.99}{d} \\
\multicolumn{5}{l}{($\varepsilon$-greedy action with $\varepsilon = 0.2$, as well as Q-learning parameters $\alpha = 0.8$ and $\gamma = 1$)} \\
\midrule
&
\textbf{\mcellt{Abnormal\\ returns:\\ mean}} &
\textbf{\mcellt{Abnormal\\ returns:\\ median}} &
\textbf{\mcellt{Volatility\\ (i.\,e. risk)}} &
\textbf{\mcellt{Mean\\ return incl.\\ transaction costs}} &
{\textbf{\mcellt{Sharpe\\ ratio}}} &
{\textbf{\mcellt{Coeff. of\\ variation}}} \\
\midrule
\textbf{Benchmarks} \\
CDAX index & {---} & {---} & \SI{0.0210}{\percent} & 0.01319 & {---} & 44.34 \\
Momentum trading & \SI{-0.0999}{\percent} & \SI{-0.0479}{\percent} & \SI{-0.0578}{\percent} & 0.04496 & 0.00037 & 96.90 \\
Portfolio trading & \SI{-0.6271}{\percent} & \SI{-0.4802}{\percent} & \SI{-0.1964}{\percent} & 0.02013 & -0.0039 & 91.65  \\
\midrule
\textbf{News trading} \\
Simple news trading & \SI{0.4197}{\percent} & \SI{0.0229}{\percent} & \SI{0.4226}{\percent} & 0.08151 & 0.0543 & 17.26 \\
Sector-specific: automobile & \SI{-0.0529}{\percent} & \SI{-0.0056}{\percent} & \SI{-0.0511}{\percent} & 0.03685 & -0.0049 & 318.62 \\
Sector-specific: chemicals & \SI{-0.0211}{\percent} & \SI{0.0000}{\percent} & \SI{-0.0278}{\percent} & 0.02712 & -0.0042 & 147.29 \\
Combined: news \& momentum & \SI{0.0335}{\percent} & \SI{-0.0331}{\percent} & \SI{0.1951}{\percent} & 0.03428 & 0.0497 & 17.13 \\
\midrule
\multicolumn{5}{l}{\textbf{Strategy learning}$^\dagger$} \\
Supervised learning & \SI{1.0948}{\percent} & \SI{0.0000}{\percent} & \SI{1.4084}{\percent} & 0.10230 & 0.1137 & 8.67 \\
Reinforcement learning & \SI{0.7708}{\percent} & \SI{0.0000}{\percent} & \SI{1.2285}{\percent} & 0.09707 & 0.1168 & 8.44 \\
\bottomrule
\multicolumn{5}{l}{$^\dagger$ Training with years 2004 and 2005; testing from years 2006 onwards}
\end{tabular}%
}%
\end{table}

While we put an emphasis on raw returns, we also provide a performance measure that incorporates simplified transaction costs. Thus, Table~\ref{tbl:comparison} reveals mean returns minus these costs. Consistent with~\cite{Hagenau.2012b,Madhavan.2002,Tetlock.2008}, we use a proportional transaction fee common in financial research -- where transaction costs are mostly varied in the range of \SI{0.1}{\percent} to \SI{0.3}{\percent}~\cite{Graf.2011} -- or assume a fixed transaction fee~\cite{Mittermayer.2004} of \US\$\,10 for buying and selling stocks. Hence, we simulate the portfolio with a simplified transaction fee for each buy/sell operation of \SI{0.1}{\percent}, equivalent to \SI{10}{bps}, in order to approximate additional costs incurred from trading operations.\footnote{A frequently used unit in finance is the basis point~(bps). Here, one unit is equal to 1/100th of \SI{1}{\percent}, \ie \SI{1}{\percent} $=$ \SI{100}{bps}.} As a result, the values might not correlate to average returns, since we perform a portfolio simulation here, where profitable trades and transaction costs are not necessarily uniformly distributed. Further information on trading costs for high-frequency orders of hedge funds can be found \eg in \cite{Edelen.2007,French.2008,Madhavan.2002}.

{Finally, Table~\ref{tbl:comparison} provides two metrics for comparing the trade-off between performance and risk. First, the coefficient of variation, measures the volatility of returns (\ie the risk) in comparison to the performance. A smaller coefficient indicates a more favorable trade-off. Interestingly, both variants of strategy learning accomplish the lowest values with \num{8.67} and \num{8.44} respectively. Second, the Sharpe ratio compares excess returns (based on the CDAX index) in comparison of the standard deviation of returns. Again, both learning approaches dominate in this metric, with reinforcement learning being slightly better. }

\subsection{Statistical tests}

This section provides statistical evidence that trading strategies can result in daily returns above zero. We perform both a one-sided Wilcoxon test and a one-sided $t$-test. In both cases, the null hypothesis tests whether a given trading strategy results in a zero mean value of daily returns. All test outcomes are provided in the form of $P$-values in \Cref{tbl:statistical_tests}. Apparently, several $P$-values are statistically significant at common significance levels. Consequently, in these cases, we can reject the null hypothesis at the corresponding significance level and conclude that the trading strategies have an above-zero mean of daily returns. This is particularly true for both simple news trading and supervised learning, providing that news trading yields statistically significant returns. {This essentially identifies methods by which news-based data can be leveraged by a decision-making model for investments.}

\begin{table}
\caption{Statistical tests to validate the hypothesis that daily returns are above zero (without transaction costs; based on strategies from \Cref{tbl:comparison}).}
\label{tbl:statistical_tests}
\centering
{\footnotesize
\singlespacing
\begin{tabular}{l SSS}
 \toprule
\textbf{Trading strategy}\hspace{0.3cm} &
{\mcellt{\textbf{Daily returns}\\ (mean)}} & 
{\mcellt{\textbf{Wilcoxon test}\\ one-sided\\ ($P$-value)}} &
{\mcellt{\textbf{$\bm{t}$-test}\\ one-sided\\ ($P$-value)}}
\\
\midrule
\textbf{Benchmarks} \\
CDAX index & \SI{0.0298}{\percent} & 0.0031\sym{**} & 0.1593 \\
Momentum trading & \SI{0.0464}{\percent} & 0.7803 & 0.3241 \\
Portfolio trading & \SI{0.0220}{\percent} & 0.0052\sym{**} & 0.3147 \\
\midrule
\textbf{News trading} \\
Simple news trading & \SI{0.4722}{\percent} & 0.0025\sym{**} & 0.0052\sym{**} \\
Sector: automobile & \SI{0.0116}{\percent} & 0.4809 & 0.4448 \\
Sector: chemicals & \SI{0.0184}{\percent} & 0.2646 & 0.3820 \\
Combination news \& momentum & \SI{0.2001}{\percent} & 0.0326\sym{*} & 0.0050\sym{**} \\
\textbf{Strategy learning} \\
Supervised learning & \SI{1.1807}{\percent} & 0.0000\sym{***} & 0.0000\sym{***} \\
Reinforcement learning & \SI{1.1506}{\percent} & 0.0000\sym{***} & 0.0000\sym{***} \\
\bottomrule
\multicolumn{4}{r}{Statistical significance levels: $^{***} 0.001$, $^{**} 0.01$, $^{*} 0.05$} 
\end{tabular}
}
\end{table}

We repeat the analysis for the case of abnormal returns. Hence, the null hypothesis tests whether a given trading strategy results in a zero mean value of daily abnormal returns. All test outcomes are provided in the form of $P$-values in \Cref{tbl:statistical_tests_abnormal_returns} (only for those with above-zero mean). Apparently, several $P$-values are statistically significant at common significance levels. Consequently, in these cases, we can reject the null hypothesis at the corresponding significance level and conclude that the trading strategies have an above-zero mean of daily abnormal returns. Here, we see such outcomes for simple news trading, supervised and reinforcement learning. All indicate that news-based data can be a viable input for an investment decision model. 

\begin{table}[H]
\caption{Selected statistical tests to validate the hypothesis that daily abnormal returns are above zero (without transaction costs; based on strategies from \Cref{tbl:comparison}).}
\label{tbl:statistical_tests_abnormal_returns}
\centering
{\footnotesize
\singlespacing
\begin{tabular}{l SSS}
 \toprule
\textbf{Trading strategy}\hspace{0.3cm} &
{\mcellt{\textbf{Daily abnormal}\\ \textbf{returns}\\ (mean)}} & 
{\mcellt{\textbf{Wilcoxon test}\\ one-sided\\ ($P$-value)}} &
{\mcellt{\textbf{$\bm{t}$-test}\\ one-sided\\ ($P$-value)}}
\\
\midrule
\textbf{News trading} \\
Simple news trading & \SI{0.4197}{\percent} & 0.0167\sym{*} & 0.0126\sym{*} \\
Combination news \& momentum & \SI{0.0335}{\percent} & 0.9387 & 0.3292 \\
\textbf{Strategy learning} \\
Supervised learning & \SI{1.0948}{\percent} & 0.0000\sym{***} & 0.0000\sym{***} \\
Reinforcement learning & \SI{0.7708}{\percent} & 0.0000\sym{***} & 0.0009\sym{***} \\
\bottomrule
\multicolumn{4}{r}{Statistical significance levels: $^{***} 0.001$, $^{**} 0.01$, $^{*} 0.05$} 
\end{tabular}
}
\end{table}

\subsection{Example}

\begin{figure}
\centering
\includegraphics[width=.9\linewidth]{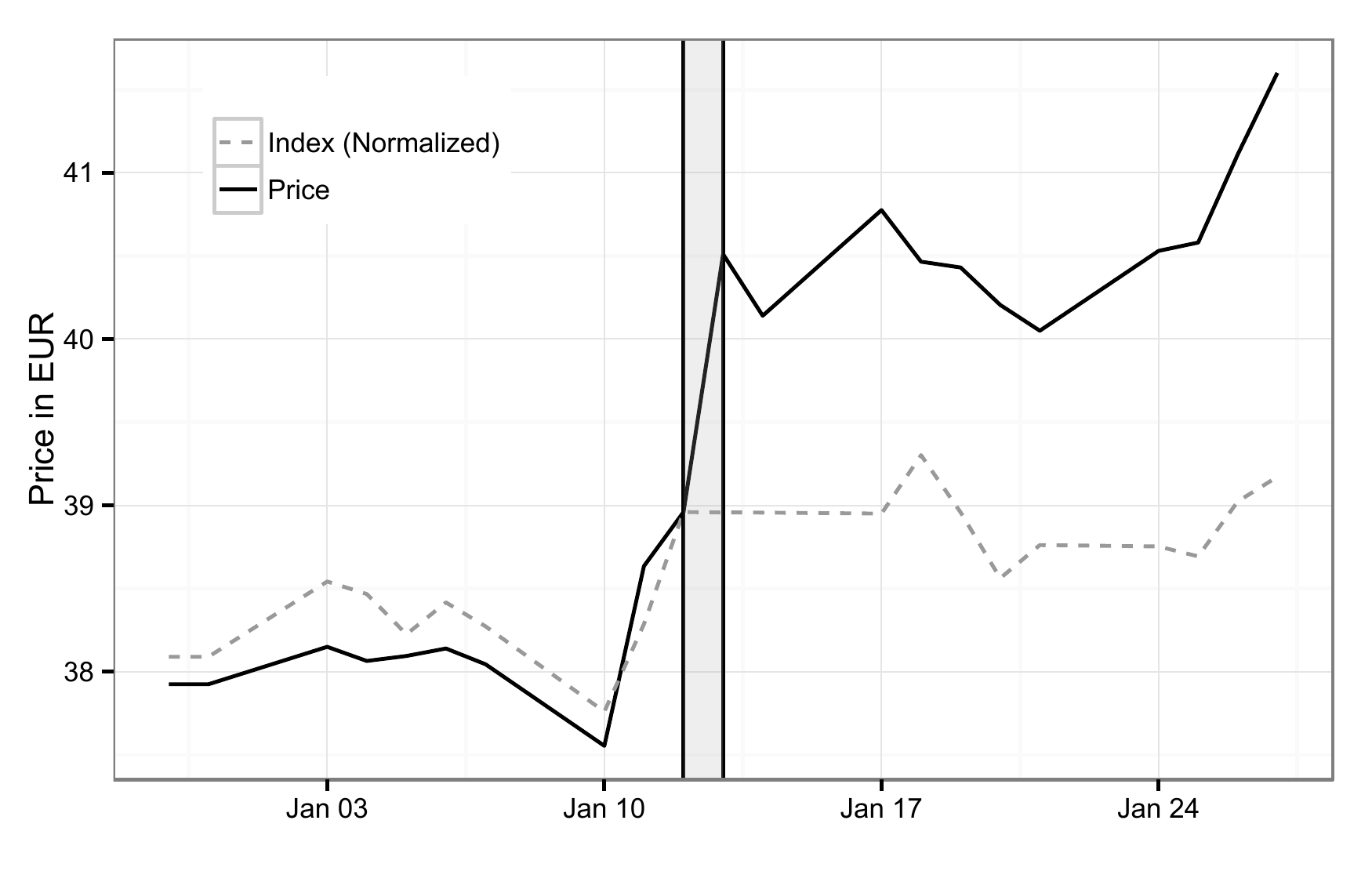}
\caption{Plot shows the price curves in an exemplary setting. The day of the ad~hoc announcement is shaded in gray background. The sample stock (\ie SAP) increases in price as a result of the news, similar to the CDAX index (which has been normalized to the price of the stock before the news disclosure).}
\label{fig:example_trade}
\end{figure}

This section demonstrates our Decision Support System in a live setup. For demonstration, we use the ad~hoc announcement titled \emph{\textquote{SAP AG: SAP Announces Record Fourth Quarter 2010 Software Revenue}}, which was released on January 13, 2011 4:44~p.\,m. by SAP AG.\footnote{Available from \url{http://www.dgap.de/dgap/News/adhoc/sap-sap-veroeffentlicht-rekordergebnis-fuer-softwareerloese-quartal/?newsID=655932}; last accessed April 18, 2016. There are a total of 11 announcements released on January~13 and the DSS might thus pick any of it or follow our evaluation above and only consider the first.} Its sentiment $S(A)$ accounts for $4.0699 \times 10^{-5}$; thus indicated an overall positive content. Depending on the current state, the system can decide between investing into the current stock, holding the previous one or following any benchmark strategy. Accordingly, we present the stock price of SAP and the CDAX index in \Cref{fig:example_trade}, while the index is normalized to match the price of the stock on the day prior to the disclosure. Apparently, the SAP stock shows a large jump in price due to the arrival of this announcement and it even outperforms the market during the subsequent days. The stock price jumps by \SI{3.98}{\percent} on the day of the disclosure, while the CDAX reference index remains fairly constant. As a result, we observe an abnormal return of \SI{3.79}{\percent}.

In addition, \Cref{tbl:example_strategies} provides a comparison of the trading decision. Both the benchmarks and rule-based trading cannot generate any profits on that disclosure day. However, both learned strategies recommend to invest in the stock and thus achieve a positive return on the disclosure day. 

\begin{table}
\caption{Comparison of different strategies based on SAP disclosure from January 13, 2011. Returns in parenthesis are based only on that specific stock ignoring interference from subsequent trades.}
\label{tbl:example_strategies}
\centering
{\scriptsize
\begin{tabular}{lSS}
\toprule
\textbf{Trading strategy} & \textbf{Return on disclosure day} & \textbf{Return across 10 days} \\
\midrule
CDAX index & \SI{0.00}{\percent} & \SI{5.51}{\percent} \\
Momentum trading & \SI{0.00}{\percent} & \SI{2.47}{\percent} \\
Portfolio trading & \SI{-0.11}{\percent} & \SI{-2.60}{\percent} \\
\midrule
Simple news trading & \SI{-1.40}{\percent} & \SI{-11.57}{\percent} \\
Combination news \& momentum & \SI{0.04}{\percent} & \SI{-10.28}{\percent} \\
\midrule
Supervised learning & \SI{3.98}{\percent} & {(\SI{5.52}{\percent})} \\
Reinforcement learning & \SI{3.98}{\percent} & {(\SI{5.52}{\percent})} \\
\bottomrule
\end{tabular}
}
\end{table}

\section{Discussion}

This section provides a discussion of managerial implications and limitations.

\subsection{Managerial implications}

This paper aims to develop the core functionality of an automated system for triggering news-based trading decisions. Although it is part of a larger algorithmic trading system, the system is often regarded as a Decision Support System~\cite{Gagnon.2013}. {In the above evaluation, such a system demonstrated that news trading can yield a profitable scenario.} In practice~\cite{Gagnon.2013}, this kind of Decision Support System may consist of 10 to 100 data/text mining algorithms. Text mining algorithms have lately contributed to the reduction of market noise from news by extracting only the relevant events from the overall news pile~\cite{GrossKlumann.2011}. As a consequence, all model parameters must frequently be calibrated to the latest datasets.

One of the challenges remaining is that of the information quality of news. Although news plays a significant role in driving stock prices, it still leaves a large portion of noise~\cite{Tetlock.2008}. This inevitably results in unpredictable and possibly reduced market reactions when signals are unclear~\cite{Tetlock.2010}. Different data sources might provide further insights into the costs of additional noise. For example, innovative news sources~\cite{Gagnon.2013,Hochreiter.2015}, such as social media like Twitter, offer a massive volume of new textual materials, which are published quickly and without quality control. Related approaches also incorporate Google query volume for search terms related to finance~\cite{Preis.2013} and Internet stock message boards~\cite{Antweiler.2004}. Each of these sources has its own advantages, but practitioners are likely to implement a combination. Regardless, we need a remedy that works reliably even when facing noisy data. To overcome this challenge, we propose news-trading strategies in the form of rule-based algorithms and strategy learning that show a low-risk component. 

The above research opens up several other research streams. The proposed Decision Support System with its trading strategies can also be highly relevant for high-frequency trading~\cite{Aldridge.2013} and provides valuable insights for this adjacent discipline, relying purely on automated trading algorithms. News trading also intersects with portfolio management problems~\cite[for example]{Mitra.2009}, in which investors make decisions about which portfolio to hold on an ex-ante basis without knowing future returns. Though these returns are unknown, market participants make decisions and execute transactions based on their expectations of the market. 

Managers on the road to implementing such news trading systems can choose between a vast number of options. However, we have seen that functional verifications are rare. It is, therefore, \emph{\textquote{no surprise that news trading remains highly specialized and continues to be based on disparate decision-making tools and analytical models}}~\cite{Gagnon.2013}.

\subsection{Generalizability and limitations}
\label{sec:limitations}

Caution needs to be exercised when transferring the previous results into practice. Therefore, we discuss limitations and generalizability in the following. Our paper aims at proposing and evaluating a Decision Support System with novel trading strategies based on machine learning in order to decide whether to buy the stock belonging to the news disclosure or invest in an alternative benchmark.

\begin{figure}
\centering
\includegraphics[width=.45\linewidth]{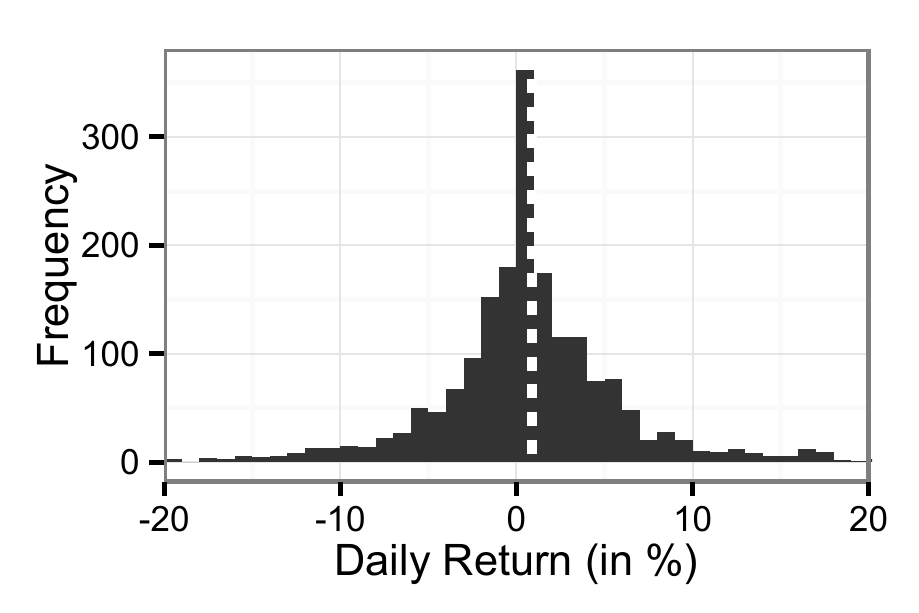}
\includegraphics[width=.45\linewidth]{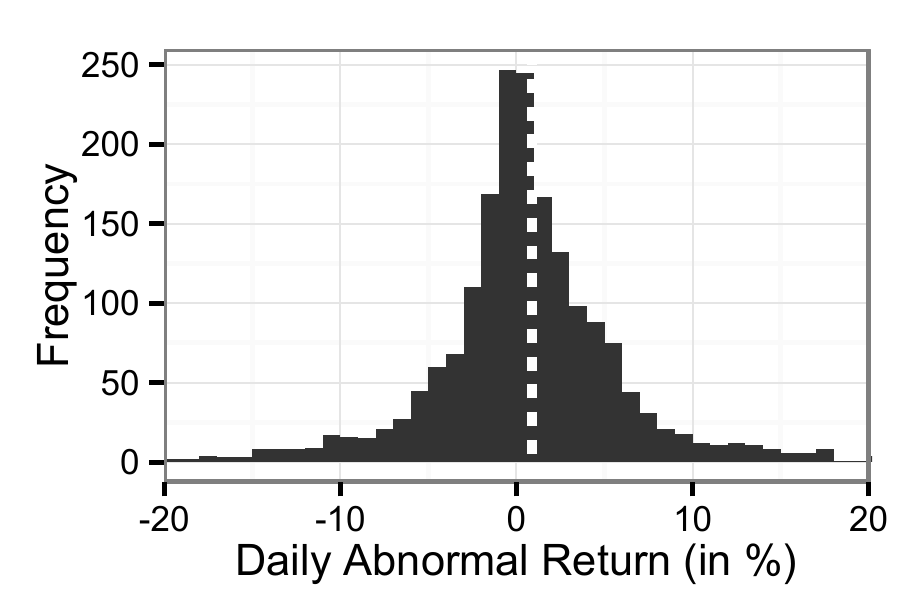}
\caption{Histogram of returns~(left) and abnormal returns~(right) on the day of the disclosure, where the vertical bars denote the corresponding mean value.}
\label{fig:histogram_news_return}
\end{figure}

Our intention is not to unravel trading signals in the first place but rather to compare strategies that convert signals into financial gains. For that reasons, we focus purely on the arrival of new information and its corresponding effect on the market. For instance, \Cref{fig:histogram_news_return} depicts stock market response to ad~hoc announcements, where new information causes non-zero average returns. With perfect prediction, one can potentially turn all positive and negative returns into financially rewarding trades. In addition, the content of ad~hoc announcements is required to be novel and relevant for the stock market. The corresponding concept is formalized in the semi-strong form of the efficient market hypothesis according to which stock prices adapt upon the disclosure of novel information \cite{Fama.1965}. Accordingly, stock prices changes will remain even when more players trade on news signals; however, it affects how profits are distributed among them. 

The Decision Support System can work in any situation where novel information enters the market and triggers a response in stock prices. Hence, one can adapt the DSS by replacing the sentiment variable with any numeric value specifying the positivity of the information. Examples can be come in the form of more macro-economic announcements (\eg regarding interest rates, unemployment rates) or firm-specific disclosures (\eg earnings calls).

In contrast, other works focus on revealing unknown variables or patterns. Such objectives are frequently termed \emph{data dredging} or \emph{data snooping}, which utilize data mining to uncover patterns in data \cite{White.2000}. These test numerous variables and combinations thereof to identify predictors that might show a correlation with future stock returns \cite{Hu.2015}. However, one needs to be careful to circumvent a \emph{look-ahead bias}, \ie using variables that were not present at the time of the trade \cite{Neuhierl.2011}. Such predictors frequently reveal only a limited return predictability. For instance, a recent study \cite{McLean.2016} tests portfolio returns for 97 variables assumed to predict cross-sectional stock returns. Examples of such variables include analyst values, bid-ask spread, turnover, investments and tax, but not the arrival of news. The findings show that returns drop once the predictor has been published in a academic journal indicating that investors adapt to mispricing and thus diminish excess profits.

Hence, the DSS shows a possible path to reaping financial benefits from the arrival of new information. However, this entails several practical requirements and challenges. First of all, the corresponding stocks need to be highly liquid in order to place larger orders (and define the order book accordingly~\cite{Obizhaeva.2013}). In addition, hedge funds need highly sophisticated analytical framework (to predict the market response based on the novel information) and the necessary IT infrastructure. The latter means high-performance servers and low-latency network connections in direct proximity to the stock exchange as automated traders currently trade on news releases and place orders within milliseconds (cf.~\cite{Foucault.2016}). For the same reasons, huge initial expenditures are necessary for new hedge funds to enter the market where automated traders already operate. Finally, competition is likely to affect how information is processed and demands constant advances regarding the technological and analytical infrastructure, thus limiting potential profit. 

\section{Conclusion and outlook}
\label{sec:conclusion}

Although it is a well-known fact that financial markets are very sensitive to the release of financial disclosures, the way in which this information is received is far from being studied sufficiently. Not until recently have researchers started to look at the content of news stories using very simple techniques to determine news sentiment. Typically, these research papers concentrate on finding a link between the qualitative content and the subsequent stock market reaction. To harness this relationship in practice, news trading combines real-time market data and sentiment analysis in order to trigger investment decisions. Interestingly, what previous approaches all have in common is that they rarely study and compare trading strategies.

As a remedy, this paper evaluates algorithmic trading strategies within a Decision Support System for news trading. We thus propose and compare different variants (such as supervised learning and reinforcement learning) for strategy learning in order to enhance news-based trading. {As a result, our Decision Support System can make profitable investment decisions by utilizing news-based data.} 
 However, the inherent prediction errors of these methods imply a certain degree of risk. Altogether, we contribute to the understanding of information processing in electronic markets and show how to enable decision support in financial markets.

This paper opens avenues for further research in two directions. First, a multi-asset strategy could be beneficial in spreading risk. To model such a strategy, intriguing approaches include Value-at-Risk~(VaR) measures, as well as techniques from portfolio optimization. Second, it is worthwhile to improve the forecast of asset returns by including a broader set of exogenous predictors. As such, possible external variables might include stock market indices, fundamentals describing the economy and additional lagged variables. Further enhancements would also result from embedding innovative news sources, such as social media. In addition, practitioners might be interested in gauging financial gains when repeating this study with intraday returns or more realistic transaction costs. Altogether, the accuracy of predicting stock return directions and triggering beneficial trading signals would be greatly improved through such refinements.

\section*{Acknowledgment}

The valuable contributions of Dirk Neumann, Laura Cuthbertson, Ryan Grabowski and Simon Schmidt are gratefully acknowledged.



\bibliographystyle{model1-num-names}
\bibliography{literature}







\end{document}